\documentclass[12pt]{article}

\usepackage{amsmath}
\usepackage{amsfonts}

\begin{document}

\title{Replica procedure for probabilistic algorithms \\ as a model of gene duplication}

\author{ A.Yu.Khrennikov, S.V.Kozyrev}

\maketitle

\begin{abstract}
In the present paper we propose to describe gene networks in biological systems using probabilistic algorithms. We describe gene duplication in the process of biological evolution using introduction of the replica procedure for probabilistic algorithms. We construct the examples of such a replica procedure for hidden Markov models. We introduce the family of hidden Markov models where the set of hidden states is a finite additive group with a  $p$-adic metric and build the replica procedure for this family of markovian models.
\end{abstract}

\section{Introduction}

Different methods of physics, in particular, probabilistic methods found application in genetics. In the present paper we introduce for applications to evolution of genomes the analogue of the replica procedure, which was used in the statistical physics of disordered systems. The replica method, cf. \cite{MPV}, \cite{Dotsenko}, was applied to description of states of disordered systems, in particular, spin glasses. In the present paper we propose to use replicas for investigation of gene duplication. One of the examples of this approach is based on the application of $p$-adic numbers and probabilistic models related to $p$-adic mathematical physics \cite{VVZ}.

One of the approaches to investigation of gene networks (genetic regulatory networks, or networks of interacting genes) and related metabolic networks in molecular biology, and also to investigation of gene regulation (regulation of gene expression) is the modeling of the mentioned networks using the corresponding system of kinetic equations. This system of kinetic equations describes metabolic reactions and levels of gene expression. An alternative approach to gene networks describes this network as a computational model which performs computations according to some algorithm (for example, a Boolean network).

In the present paper we propose to describe a gene network as a probabilistic algorithm. Let us recall that a probabilistic (or randomized) algorithm differs from the standard (deterministic) algorithm as follows: probabilistic algorithm performs commands with some probability. Therefore a probabilistic algorithm depends on the set of parameters (probabilities). One can put in cor\-res\-pon\-dence to a probabilistic algorithm a system of kinetic equations which describes the rate of operation of some commands of the algorithm. This allows to unify the kinetic and the algorithmic descriptions of a gene network.

In this approach the parameters of a probabilistic algorithm correspond to the levels of gene expression. For discussion of theory of algorithms see \cite{Kolmogorov}, for introduction to probabilistic algorithms cf. \cite{Kitaev}. Probabilistic and quantum algorithms with applications to some problems of bioinformatics, in particular, to sequence alignment were discussed in \cite{OhyaVolovich}. For discussion of analysis of genomes cf.  \cite{Koonin} and for review of gene networks cf. \cite{KooninWolfKarev}.

One of the motivations for application of probabilistic algorithms to gene networks is the evolvability of genomes. According to the theory of neutral evolution \cite{Kimura} the majority of mutations (changes of a genome in the process of biological evolution) does not influate the fitness of the corresponding organisms. From the point of view of a genome as an algorithm this is not natural --- random transformations of a program will break this program, i.e. will transform an efficient algorithm to inefficient. Here efficient algorithm is an algorithm which is able to perform the needed computations, in application to biology this will correspond to the genome of a biologically fit organism.

For a probabilistic algorithm continuous transformations of the parameters of the algorithm are possible. In application to gene networks these transformations correspond to variation of levels of gene expression. These transformations could be achieved by substitutions in regulatory sequences of the genome.

Transformations of a genome in the process of evolution are not restricted to variation of levels of gene expression. One of the important mechanisms of evolution is the gene duplication, cf. \cite{Ohno}. Under the gene duplication some parts of the genome (for example, the whole genome) can be duplicated several times (i.e. the new genome will contain several copies of the part of the old genome). Sequences which are the duplicates of some sequence are called paralogous. After the duplication the different copies of a gene may work as the initial gene, may be switched off, and may evolve obtaining new functions (the process duplication--specialization). The horizontal gene transfer can be considered as a particular case of gene duplication in the union of genomes of the different organisms.

In order to construct the model of biological evolution one has to describe the class of probabilistic algorithms which correspond to gene networks and the family of transformations which describe point mutations and gene duplication. These transformations, according to the theory of neutral evolution, should transform an efficient algorithm to the efficient algorithm with high probability. In the present paper we consider the model of gene duplication for hidden Markov models.

We discuss the analogy between the phenomenon of gene duplication and the replica procedure which was applied in the statistical physics of complex systems (in particular, the replica method in the theory of spin glasses). The replica procedure transforms the Hamiltonian of a complex system to several copies (replicas) of this Hamiltonian. The analogous transformation is applied to the observables. The quenched state for the complex system in the framework of the replica approach is computed as a result of interaction of several replicas of this system, cf. \cite{MPV}, \cite{Dotsenko}.

Analogously the gene duplication substitutes a part of a genome by several copies of this part. In the genetic program these parts of the initial genome will work parallelly. If we consider the genetic program as a probabilistic algorithm then the gene duplication looks like a replica procedure for this algorithm. We arrive to the following problem  --- {\it to introduce a natural definition of the replica procedure for probabilistic algorithms}. This procedure should transform an efficient algorithm to an efficient algorithms with nonzero probability. The different replicas will correspond to paralogous sequences (genes or regulatory sequences).

Let us note that for a general probabilistic algorithm there is no natural definition of a replica procedure (and, if this procedure exists, it should not be unique). We also do not claim that we build in this paper a realistic model of gene regulation for some existing gene network, our aim here is to describe some nontrivial examples of replica procedure for probabilistic algorithms.

In the present paper we consider some examples of replica procedures for hidden Markov models, or HMM (some simple class of probabilistic algorithms). In particular, we build an example of a hidden Markov model where the set of hidden states is the additive finite group with the $p$-adic metric and introduce the replica procedure for this model. Let us note that the  $p$-adic methods of description of the genetic code were developed in  \cite{genetic_code,PAM,DD,Andr1,Rumer}.

\section{Replica procedure for HMM}

In the present section we introduce the definition of the replica procedure for hidden Markov models.

A hidden Markov model (or HMM) $F(f(t))$ is a random function of a Markov chain. Here $f$ is a Markov chain with the discrete time (i.e. the time $t$ is a natural number) which takes values in the finite set $X$ (the set of hidden states of the model), $F:X\to Y$ is a random map from the finite set $X$ of hidden states to the finite set $Y$ of output (or production, or emission) states of the model. The map $F$ describes the family of emission probabilities of the HMM.

Therefore a hidden Markov model is described by the maps
$$
{~\atop\mathbb{N}}{f\atop\longrightarrow}{~\atop X} {F\atop\longrightarrow} {~\atop Y}
$$
where the Markov chain $f$ is defined by the set of transition probabilities $p_{xx'}:x\to x'$, $x,x'\in X$.

The introduced in the present paper replica procedure for hidden Markov models is defined with the help of the replacement of the set $X$ of hidden states of the model by the direct product $X\times R$, where $R$ is a finite set (the set of replicas). 

We discuss the following biological interpretation. A hidden Markov model generates a biological sequence (for example, a DNA sequence). The set $Y$ is a set of possible elements of the mentioned biological sequence (for a DNA this will be the set of nucleotides $\{A,T,G,C\}$), the set $X$ is the set which describes the different regimes of generation of sequences. The replica procedure is a model of gene duplication  --- the different replicas correspond to the different paralogous sequences (genes or regulatory sequences).

Let us describe the hidden Markov model for the replica symmetric case. The replica symmetry in our approach corresponds to the gene duplication for the case of neutral evolution. Neutral evolution transforms a probabilistic algorithm to an equivalent probabilistic algorithm (i.e. the algorithm which generates the same results with the same probabilities).
This kind of hidden Markov model will be described by the composition of maps
$$
{~\atop\mathbb{N}}{\widetilde{f}\atop\longrightarrow}{~\atop X\times R} {\widetilde{F}\atop\longrightarrow} {~\atop Y}
$$
where $\widetilde{f}$ is a Markov chain with the set of transition probabilities
$$
p_{(x,r);(x',r')}=p_{xx'}(\delta_{rr'}+c(1-\delta_{rr'})),\qquad c\in [0,1].
$$
The transition between the different replicas $r$, $r'$ will have the probability which is proportional to the coefficient $c$.

The map $\widetilde{F}$ for the replica symmetric case will be given by the formula
$$
\widetilde{F}(x,r)=F(x),
$$
i.e. this random map will not depend on the replica index $r$ (a copy of a gene in the set of paralogs). Therefore the replica symmetric hidden Markov model $\widetilde{F}(\widetilde{f}(t))$ is equivalent to the initial hidden Markov model $F(f(t))$. This means that $\widetilde{F}(\widetilde{f}(t))$ generates the same (statistically) sequences of elements of $Y$ as $F(f(t))$. In particular the described replica procedure will map an efficient HMM to an efficient HMM.

In general one can consider a replica procedure with broken replica symmetry. In this case the map $\widetilde{F}$ will depend on the replica index $r$. Models with broken replica symmetry will describe the specialization of genes after duplication. A hidden Markov model with the broken replica symmetry will not be equivalent to the initial hidden Markov model.

\section{The $p$-adic HMM}

In the present section we discuss hidden Markov models where the sets of hidden states are hierarchical (i.e. are described by some ultrametric spaces). The simplest example of a hierarchical Markov model has the form
$$
{~\atop\mathbb{N}}{f\atop\longrightarrow}{~\atop X} {F\atop\longrightarrow} {~\atop Y}
$$
where:

1) The set $Y$ of output (or production, or emission) states of the Markov model is a finite set which in the model under consideration is taken to be equal to the set of nucleotides. We consider the 2-adic parametrization of the set $Y=\{A,U,G,C\}$ introduced in \cite{genetic_code}, \cite{PAM}, i.e. the parametrization of $Y$ by the space $\mathbb{F}_2^2$ (2--dimensional space over the field of residues modulo 2). In the 2-adic approach the nucleotides are parametrized by the pairs of 0 and 1 as follows
$$
\begin{array}{|c|c|}\hline
A & G \cr\hline U & C \cr\hline
\end{array}=\begin{array}{|c|c|}\hline 00 & 01
\cr\hline 10 & 11 \cr\hline
\end{array}
$$

2) The set $X$ of hidden states of the Markov model is an ultrametric space. In the example under consideration $X=\mathbb{Z}/2^{N}\mathbb{Z}$, $N>0$, i.e. the set $X$ is an additive group of residues modulo $2^{N}$ with the naturally defined 2-adic metric. We consider the Haar measure on the group $X$, where the measure is normalized in such a way that the measure of the group $X$ is equal to one.

3) The map $f$ is a Markov chain taking values in the set $X$ of hidden states. The family of transition probabilities of this Markov chain describes the discrete $2$-adic diffusion, i.e. has the form
\begin{equation}\label{q}
p_{xy}=q(|x-y|_2),\qquad q(\cdot)>0,\qquad  \int_{x\ne 0} q(|x|_2)dx<1,
\end{equation}
(where the integral is taken with respect to the mentioned Haar measure).

4) The random map $F:X\to Y$ is constructed as follows. We put in correspondence to any ball $J\subset X$ (including balls of zero diameter, i.e. points) the characteristic function $\chi_J$ of this ball. This function takes values in $\mathbb{F}_2$ (i.e. is equal to one in the ball and to zero outside the ball, where one and zero are considered as elements of $\mathbb{F}_2$). Also we put in correspondence to a ball $J$ the random variable $\phi_J$ taking values in $\mathbb{F}_2$, where $\phi_J$  is equal to 1 with the probability $p_J$ which depends on the ball $J$, and this probability is a monotonously increasing function of a ball (i.e. for $I\supset J$ one has $p_I>p_J$)\footnote{For example, one can put $p_J$ to be proportional to the Haar measure of the ball $J$.}. Let also the random variables $\phi_J$ for the different $J$ be independent.

Let us consider for the point $x\in X$ the maximal increasing sequence $\{J\}$ of balls which contain $x$ (i.e. the minimal ball in this sequence is $x$ and the maximal ball is $X$). The map $F_1: X\to \mathbb{F}_2$ puts in correspondence to a point $x\in X$ the random element of $\mathbb{F}_2$ which is constructed as follows:
\begin{equation}\label{F_1}
F_1(x)=\sum_{J} \phi_J\chi_J(x).
\end{equation}
The summation runs over the sequence of balls $\{J\}$, $x\in J$.

The map $F:X\to \mathbb{F}_2^2$ is constructed as the sum of the two independent maps $F_1$ acting at each of the coordinates in $\mathbb{F}_2^2$.

\bigskip

Let us discuss the example of the replica procedure for the introduced in the present section hierarchical hidden Markov model. The replication of the set $X$ of hidden states of the model (transition from $X$ to $\widetilde{X}$) in this case can be considered as related to the map of taking ${\rm mod}\, 2^N$ residue (i.e. the projection)
$$
\widetilde{X}\to X:\qquad \mathbb{Z}/2^{M}\mathbb{Z}\to \mathbb{Z}/2^{N}\mathbb{Z},\qquad M>N.
$$

Therefore the set $\widetilde{X}$ differs from the set $X$ at small distances, i.e. points of $X$ correspond to balls in $\widetilde{X}$ with the diameter $2^{-M}$. Each of these balls contain $2^{M-N}$ points. As a set this space is in one to one correspondence with the direct product of $X$ and the set consisting of $2^{M-N}$ elements (this allows to compare the above definition of $\widetilde{X}$ and the definition of the replica procedure for hidden Markov models in the previous section).

The map $\widetilde{f}$ is constructed by extension of the transition probability $q(\cdot)$ of the Markov chain $f$ to small distances, i.e.  $\widetilde{f}$ corresponds to the transition probability $\widetilde{q}(\cdot)$ where $\widetilde{q}(x)=q(x)$ for $|x|_2>2^{-M}$ and for $|x|_2\le 2^{-M}$ the transition probability $\widetilde{q}(\cdot)$ is defined in some arbitrary way (taking into account the conditions mentioned in the formula (\ref{q})).

Analogously, the map $\widetilde{F}$ is built from the map $F$ by extension to smaller distances. We extend the set of independent random variables $\{\phi_J\}$ by random variables corresponding to balls with the diameters satisfying $|J|_2\le 2^{-M}$ (taking into account the mentioned above properties of the set $\{\phi_J\}$). We define the map $\widetilde{F}_1$ by the formula (\ref{F_1}) (which now contains contributions from smaller balls), and define the map $\widetilde{F}$ as a pair of independent $\widetilde{F}_1$.

We have constructed the natural replica procedure for the described example of hierarchical hidden Markov model. Let us note that the introduced in the present section replica procedure in general differs from the replica symmetric case considered in the previous section, since the maps $\widetilde{f}$ and $\widetilde{F}$ defined in this section are not necessarily coincide with the replica symmetric maps of the previous section.

\bigskip

\noindent{\bf Acknowledgments}\qquad This paper was partially supported by the grant ''Mathematical Modeling'' of the Linnaeus University (V\"axj\"o, Sweden). One of the authors (S.K) gratefully
acknowledge being partially supported by the grant of
the Russian Foundation for Basic Research
RFFI 11-01-00828-a, by the grant of the President of Russian
Federation for the support of scientific schools NSh-7675.2010.1 and
by the Program of the Department of Mathematics of the Russian
Academy of Science ''Modern problems of theoretical mathematics''.

\end{document}